\documentstyle{l-aa}

\begin{document}

\thesaurus{02(03.12.1)}

\title{On the preferred length scale in the anisotropy field
of extragalactic IRAS sources}
\author{R.~Fabbri\inst{1} \and V.~Natale \inst{2}}

\offprints{R.Fabbri}
\institute{Dipartimento di Fisica dell'Universit\'{a},
 Sezione di Fisica Superiore,
 Via S. Marta 3,  I-50139 Firenze,  Italy
 \and CAISMI-CNR,  Largo Fermi 5,  I-50125 Firenze,  Italy}

\date{Received ; accepted }

\maketitle
\begin{abstract}
We investigate the existence and meaning of a preferred length in the
large-scale distribution of 60-$\mu $m IRAS sources applying three tests to
the 2-dimensional distribution on the celestial sphere, involving
respectively the rms fluctuation and the correlation function of the source
number density, and the peak number statistics. For HDM models we find a
best value $\lambda _0=$ $45-55\;h^{-1}$ Mpc for the peak of the spectral
function $P(k)$, close to the preferred wavelength $\lambda =30-40\;h^{-1}$
Mpc of na\"\i ve single-scale perturbation models. A feature should indeed
be generated in this range by the anticorrelation region of $\xi (r)$,
although it is not detected by recent Fourier transform analyses of redshift
catalogs. Our tests provide acceptable fits for CDM models with either $%
\Omega _0h=0.3-0.5$ or $\Omega _0h=0.1-0.2$, in agreement with results in
current literature. However, the quality of such fits is less satisfactory
than for HDM models, with the exception of the peak number statistics, and
we are not able to find a parameter interval satisfying all of our tests
simultaneously. These results may indicate that the cosmic spectrum is more
complicated than the smooth shapes predicted by current theoretical models.
\end{abstract}

\section{Introduction}

The anisotropy field in the 2-dimensional distribution of IRAS sources on
angular scales of $1^{\circ }-70^{\circ }$ was recently used to derive a
characteristic length scale of about $50h^{-1\;}$Mpc (Fabbri \& Natale 1993;
the Hubble constant $H_0\;$is here $100h\;$km s$^{-1}$Mpc$^{-1}$),
significantly larger than the dipole-dominating scale (Villumsen \& Strauss
1987; Clowes et al. 1987) and comparable to the distance $D_0\approx
30h^{-1} $ Mpc at which the galaxy-galaxy correlation function $\xi (r)$
becomes negative (Guzzo et al. 1991; see also Davis \& Peebles 1983). Torres
et al. (1994) have shown that the Fourier transform of $\xi (r)$, suitably
modeled to reproduce the data of Guzzo et al., should induce a peak in the
spectral function $P(k)\;$of the inhomogeneities at a wavelength $\lambda
_{{\rm peak}}\simeq 1.5D_0$; further, they find that best fitting on COBE
data, involving the angular correlation function of the cosmic background
radiation, provides $\lambda _{{\rm peak}}$ $=51\pm 18\;h^{-1}$ Mpc. We are
thereby lead to suspect that the characteristic length in IRAS may be close
to a peak of $P(k)$, although not necessarily to its absolute maximum. This
interpretation, even in its weakest form, seems to be not consistent with
several works inferring $P(k)$ from the 3-dimensional distributions of
visible and IR galaxies and clusters (Vogeley et al. 1992; Fisher et al.
1993; Einasto et al. 1993; Jing \& Valdarnini, 1993). These works, based on
Fourier transforms of red\-shift-space distributions, provided spectra
which, although not completely consistent with each other, steadily increase
up to or beyond $100h^{-1}$ Mpc: When some peak is detected, it is located
at large wavelengths, in the range $180-300\;h^{-1}$ Mpc. However, Bahcall
et al. (1993) argue that large-scale velocities induced by the inhomogeneity
field itself, not properly taken into account in the above works, may lead
to largely overestimate the large-scale portion of the spectrum in real
(rather than redshift) space. Corrections are strongly model dependent, and
for some of the CDM models considered by Bahcall et al. they shift the peak
wavelengths in the IRAS and CfA spectra down to $\simeq 100h^{-1}$ Mpc. Is is
thereby important in our view to exploit alternative techniques that are not
affected by the problem of large scale velocities.

In this paper we reconsider the question of the significance and reliability
of the preferred scale that we identified in Version I of the IRAS Point
Source Catalog. We use a larger sample including 12511 sources derived from
Version II of the Catalog, and we perform three statistical tests based on
finite-size bins. These include the shapes of the functions (i) $\Delta N_{%
{\rm rms}}(\gamma ,\gamma )$, the rms fluctuation in the source number
density vs. angular scale $\gamma \;$in the range $1^{\circ }-72^{\circ }$,
(ii) $C(\gamma ,\sigma )$, the autocorrelation function of the number
density at$\;2^{\circ }-25^{\circ }$, and (iii) $N_{{\rm peak}}(\sigma ,T)$,
the number of local maxima in the source distribution vs. Gaussian beam
dispersion $\sigma \;$(in the range $1^{\circ }-10^{\circ }$) and
peak-height threshold $T$. The experimental data are tested against a large
number (about 2000) of theoretical models.

First we model the inhomogeneity field by isotropically stochastic
perturbations endowed with a single wavelength $\lambda $, and search for
the best values of $\lambda $, the cosmological density parameter $\Omega _0$%
, and the product $bV$ (the biasing factor of IRAS sources times the local
velocity). Although the present sample allows us detecting some dependency
of $\lambda$ on the chosen angular range (in particular using the statistics on
$N_{%
{\rm peak}}$), nevertheless we can identify a preferred range around $%
30-40\;h^{-1}\;$Mpc, irrespective of $\Omega _0$ which cannot be determined
by the present tests. We then examine continuous spectra modeling both hot
and cold matter with the usual requirement that$\;P(k)\propto k$ at large
wavelengths. We find that the HDM model in the $\Omega _0=1$ cosmology gives
excellent fits for all of our tests, except for one data point of $\Delta N_{%
{\rm rms}}$ at the largest scale $71.7^{\circ }$. The HDM peak wavelength $%
\lambda _0$ is only slightly larger than $\lambda \;$found for single-scale
models. Individual tests can give best values as high as $60h^{-1}$ Mpc with
large error bars, but two joint tests combining the peak number data with
rms fluctuation and correlation function data, respectively, give $\lambda
_0=52.5_{-7}^{+22}\, h^{-1}$ Mpc and $45.0_{-1.5}^{+5}\;h^{-1}$ Mpc. Although
we cannot derive stringent limits on $bV$ from individual HDM tests, when we
require to simultaneously satisfy all of our tests we get a preferred value
of $450-550$ km/s (in close agreement with results that we find from
single-scale models for $\Omega _0=1$).

Basically the same picture arises for other sharply-peaked spectral forms,
for which the only important parameter is the location $\lambda _0$ of the
peak of$\;P(k)$. Thus similar results for $\lambda _0\;$are found for simple
two-power spectra with slopes around $-1.5$ at large $k$. The investigation
of CDM models on the other hand provides a quite different picture. For
individual tests we find best-fit values of the shape parameter $\Omega _0h$
in one of the ranges $0.3-0.5$ or $0.1-0.2$. Low-density models are favoured
by several studies (e.g., Efstathiou et al. 1992). Now for $\Omega _0h\leq
0.5$ the peak wavelength is $\ga100h^{-1}$ Mpc; in fact, single-scale models
are quite unable to represent so broad spectra as those implied by CDM
models. However, the quality of CDM fits is less satisfactory than for HDM
models, with the exception of the peak number statistics. In addition, no
range of $\Omega _0h$ is completely satisfactory for all of our tests; this
latter fact recalls the results of Scharf and Lahav (1993). It is well known
that pure CDM models do not generally provide excellent fits to available
data, a fact which motivates studies involving mixed matter (Davis et al.
1992, Taylor \& Rowan-Roinson 1992) or a cosmological constant (Efstathiou
et al. 1992). However, the comparison of HDM and CDM fits in our work (Sect.
4) is rather impressive.

The spectral window functions that can be suitably defined for $\Delta N_{%
{\rm rms}}(\gamma ,\gamma )$ and $C(\gamma ,\sigma )$ show that such tests
weight small scales less than other 2-dim\-ensional tests from other authors
(in particular, Scharf and Lahav 1993). This provides support to the idea
that the preferred scale arising from sharply peaked spectra should
correspond to something of physical. It is also clear that such a feature
should be regarded as generated by the behaviour of $\xi (r)$ around $%
30h^{-1}$ Mpc measured by Guzzo et al. (1991). Since we do not directly
construct an explicit spectral form from data, we do not claim that our
tests on the distribution of IRAS sources necessarily imply an absolute
maximum of $P(k)\;$at scale $\lambda _0$, nor can we claim that HDM models
satisfy the bulk of available astronomical data. However, our results make
us believe that the shape of $P(k)$ below $100h^{-1}$ Mpc should not be so
simple and featureless as generally believed, and that a positive slope up
to or beyond $180h^{-1}$ Mpc is probably incorrect. According to recent
measurements of the angular correlation function in the APM survey the
spectrum rises above a power law around $30h^{-1}$ Mpc and seems to be
peaked at $100h^{-1}$ Mpc (Baugh \& Efstathiou 1993), a scale smaller than
declared by most authors on the basis of redshift surveys. In our view the
problem of the detailed shape of $P(k)$ is not so clearly solved and
deserves further investigations.

\section{Modelling}

Our models assume that the large-scale inhomogeneities are described by a
isotropic, Gaussian field of perturbations. The theory of source-count
anisotropies has been developed (Fabbri 1988, 1992) for a Friedmann
background with an arbitrary density parameter $\Omega _0$, perturbed by a
cosmic density-wave spectrum $P(k)$ such that

\begin{equation}
\label{deltarho}\left\langle \left| \frac{\Delta \rho }\rho \right|
^2\right\rangle =\int {\rm d}kk^2P(k).
\end{equation}
Here the dimension\-less wavenumber $k$ is connected to the physical
wavelength by
\begin{equation}
\label{kappa}k=\frac{2\pi c}{H_0S(\Omega _0)\lambda },
\end{equation}
with $S(\Omega _0)=\left| 1-\Omega _0\right| ^{\frac 12}$ for $\Omega _0\neq
1$ and $S(1)=\frac 12$. All the quantities that will be evaluated
theoretically are derived from the squared coefficients of a harmonic
expansion of the source-count anisotropy field $\Delta N=\left\langle
N\right\rangle \sum_{\ell m}a_{\ell m}Y_{\ell m}$. More precisely, we
consider the expectation values
\begin{equation}
\label{al}a_\ell ^2\equiv \sum_m\left\langle \left| a_{\ell m}\right|
^2\right\rangle =a_{S\ell }^2+a_{N\ell }^2,
\end{equation}
where we sum up the cosmic-structure contribution$\;a_{S\ell }^2$ given by
\begin{equation}
\label{cosmical}a_{S\ell }^2=4\pi \left( 2\ell +1\right) \int {\rm d}k\frac{%
k^2P(k)}{\left( k^2-4K\right) ^2}A_\ell ^2(k)
\end{equation}
and the noise term $a_{N\ell }^2$ originated by discrete sources
\begin{equation}
\label{noiseal}a_{N\ell }^2=\frac{\left( 2\ell +1\right) }{\left\langle
N\right\rangle }.
\end{equation}
In Eq. (\ref{cosmical}) $K$ denotes the sign of space curvature and the
functions $A_\ell (k)$ are suitable coefficients introduced by Fabbri
(1988). We could also define window functions $\Psi _\ell (k)$ as in Scharf
and Lahav (1993) by setting $A_\ell (k)\propto b(k^2-4K)\Psi _\ell (k)$,
with $b\;$the biasing factor of IRAS sources.

The analysis of $A_\ell (k)$
(Fabbri 1992) shows that the source properties affect the calculation
through the biasing factor, the luminosity function, the number evolution
factor and, to a very small extent, the spectral index. We shall provide
explicit results for the exponential-logarithmic luminosity function of
Saunders et al. (1990), which best fits data with a number evolution factor $%
(1+z)^{3.7}$. However, using the double-power functions of Lawrence et al.
(1986) or Villumsen and Strauss (1987) does not change the conclusions of
Sect. 4 (adopting no source-number evolution of course, since they were
obtained with this assumption). Since in any case the assumed source
distribution extends farther than probed in redshift surveys (Strauss et al.
1990), we checked the influence of a cutoff placed at some $z_{\max }$.
Figure 1 displays some typical results for $a_\ell ^2$, showing that we get
no significant differences provided $z_{\max }\ga0.1$. A sharp cutoff at
redshifts $z_{\max }\;\la0.03$ would decrease the amplitude of high order
harmonics relative to low order ones. Thus we just assume that sources still
exist at a distance (along the light cone) equal to roughly three times the
median depth of PSC. We do not regard this as an onerous requirement: In
practice we only assume that the median depth of the catalog is determined
by the source luminosity function and the lower flux limit. The global
properties of the universe are involved through $\Omega _0$. Further, it was
shown that, because of the presence of the discrete source term in Eq. (\ref
{al}), the shape of the harmonic spectrum of anisotropies depends both on
the shape of $P(k)$ and on the product $bV$, with $V\;$the local velocity
with respect to the cosmic background radiation. Since $V\;$is generated by
the perturbation field itself, it is also a measure of the absolute
normalization of $P(k)$. (This implies that the shape of $a_\ell ^2$ vs. $%
\ell \;$should be sufficient to fix its own normalization.) However, $V$ may
be affected by nonlinear disturbances. On the other hand, the absolute
normalization of $a_\ell ^2$ is affected by high redshift sources more than
the shape of the harmonic spectrum (Fabbri 1988), and it might introduce a
systematic error on the normalization of $P(k)$. We thereby prefer to fit $%
bV\;$and the normalization factor of $P(k)$ as two independent parameters.
We then check a posteriori whether we obtain consistent results for $b\sigma
_8$, the biased fluctuation in $8h^{-1}\;$Mpc spheres, computed from these
two parameters when they are both available. For the peak number test, only
the first method (i.e., derivation from $bV$) can be used.
\begin{figure}
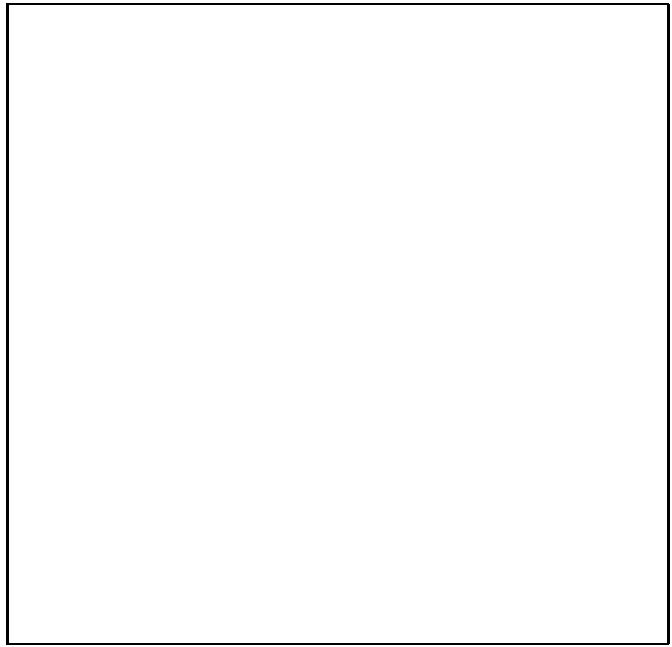

\picplace{8.5cm}
\caption[]{The squared harmonic amplitudes vs. $l$ for single-scale models with
$%
\Omega _0 = 1$ and $h\lambda =10$ and 90 Mpc. Full lines represent
overlapping curves with $z_{{\rm max}}=10^2$, 1 and 0.3. Dashed lines refer
to $z_{{\rm max}}=0.1$, and the dotted one to $z_{{\rm max}}=0.03$}
\label{f1}
\end{figure}
\begin{figure}
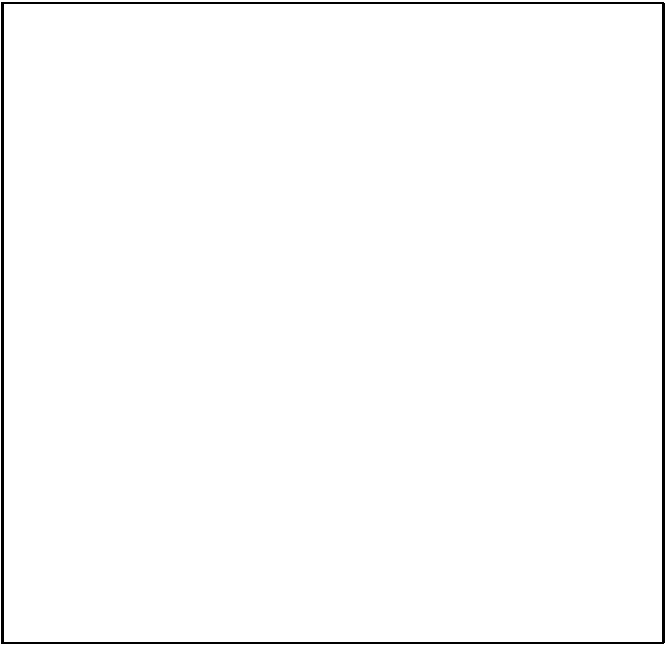

\picplace{8.5cm}
\caption[]{The window functions
 $\left| \Psi _F\right| ^2$ (full lines) and $%
\left| \Psi _C\right| ^2$ (dashed) for $\Omega _0 = 1$ and sharp-edge
beams, compared to some single-harmonic window functions of Scharf and Lahav
(1993) (dash-dotted) and to a few forms of $k^3P(k)$ (dotted lines).
For $\left| \Psi _C\right| ^2$ the
curves refer to $\gamma = 2.24^{\circ }$ and $17.9^{\circ }$;
for $\left| \Psi
_F\right| ^2$ the curve $\gamma =71.7^{\circ }$ is added. In both cases
the lowest curves correspond to largest values of $\gamma \;$for the
(arbitrarily chosen) relative normalization. The single-harmonic functions
refer to $l=4$, 10 and 40, and are computed for the 'optimal weighting
scheme' with $\Omega _0 = 0.5$, as described by Scharf and Lahav. The
spectra of HDM models (labelled by H45 and H52) correspond to $h\lambda _0=45
$ and 52.5 Mpc respectively, and those of CDM models (labelled by C50 and
C15) to $\Omega _0h = 0.50$ and 0.15}
\label{f2}
\end{figure}

As to the shape of $P(k)$, here we consider both single-scale models and
continuous spectra. For the reasons outlined above, single-scale models only
depend on three important parameters, i.e. $\lambda $, $\Omega _0$ and $bV$.
For a full investigation of their influence we generated a regularly spaced
'lattice' of 900 models for the adopted optimal set of source parameters
[i.e., using the luminosity function of Saunders et al. (1990) and a mean
spectral index of $-1.395$]; including also the luminosity functions of
Lawrence et al. (1986) and Villumsen and Strauss (1987), we investigated
about 1500 single-scale models. Continuous spectra, which were investigated
only in conjunction with the exp\-on\-ential-logarithmic luminosity
function, were modeled by simple analytic forms for hot and cold matter. HDM
models are described by
\begin{equation}
\label{hot}P(k)\propto k\exp \left[ -2\left( \frac k{k_{{\rm c}}}\right)
^{\frac 32}\right] ,
\end{equation}
with $k_{{\rm c}}$ the cutoff wavenumber. The peak wavelength is given by $%
\lambda _0=2\cdot 3^{\frac 23}\pi c\;\left[ H_0S(\Omega _0)k_{{\rm c}%
}\right] ^{-1}$. The number of models computed with the spectrum in Eq. (\ref
{hot}) was about 200, and we limited ourselves to the critical density
cosmology. We also considered about 200 models with power-law behaviours at
large $k$. We analyzed about 200 CDM models with
\begin{equation}
\label{cold}P(k)\propto k\left[ 1+1.7\widetilde{k}+9.0\widetilde{k}^{\frac
32}+1.0\widetilde{k}^2\right] ^{-2},
\end{equation}
where $\widetilde{k}=S(\Omega _0)k/(3000\cdot \Omega _0h)$. In this case the
relevant parameter is $\Omega _0h$, that we let vary in the range $(0.1-1)$.
The corresponding peak wavelength is given by $h\lambda _0\simeq 52(\Omega
_0h)^{-1}$ Mpc. All of the above continuous spectra are so adjusted as to
reproduce the standard Harrison-Zeldovich slope at large scales and exhibit
a maximum. For sharply peaked spectra, which include HDM (but not CDM)
models, only the location of this maximum is really important. Thus the
small-scale behaviour in Eq. (\ref{hot}), theoretically predicted by simple
modeling of dissipative processes in hot matter, can be easily changed
without affecting our conclusions. (This was tested investigating the
sharply-peaked two-power spectra.) Although one can argue that Eq. (\ref{hot}%
) cannot reproduce such astronomical data as the galaxy correlation function
on scales of tens of Megaparsec (see e.g. Guzzo et al. 1991), it is the
simplest one can imagine and turns out to pass our tests quite well. Also,
for the best fitted parameters arising from the tests it gives reasonable
values of $b\sigma _8$ .

For each model we calculated 300 harmonic amplitudes, in order to provide
reliable results down to small angular scales. We then derived both the
squared fluctuation in the source number per unit solid angle (for two-beam
experiments) and the angular correlation function smeared over finite
beamwidths, which are respectively given by the equations
\begin{equation}
\label{rms}
\frac{\Delta N_{{\rm rms}}^2(\gamma ,\sigma )}{\left\langle
N\right\rangle ^2}=\frac 1{2\pi }\sum_\ell a_\ell ^2F_\ell ^2(\sigma )\left[
1-P_\ell (\cos \gamma )\right]
\end{equation}
and
\begin{equation}
\label{c}
C(\gamma ,\sigma )=\frac 1{4\pi }\sum_\ell a_\ell ^2F_\ell
^2(\sigma )P_\ell (\cos \gamma ),
\end{equation}
with $\gamma $ the angular separation of beam centers (beamthrow), $\sigma %
$ the effective beamwidth, and $P_\ell $ the Legendre polynomials. The
coefficients $F_\ell (\sigma )$ are affected by the beam angular response
and are well known for both Gaussian and sharp-edge beams. In this paper any
theoretical results based on Eqs. (\ref{rms}) and (\ref{c}) refer to
sharp-edge circular beams, for which $\sigma \;$denotes the angular
diameter. Equation (\ref{rms}) was used to evaluate $\Delta N_{{\rm rms}}$
with $\sigma =\gamma $, while using (\ref{c}) $\sigma \;$is kept fixed. In
both cases we checked that calculations are accurate down to $\gamma \simeq
1^{\circ }$, while at $\gamma \simeq 0.5^{\circ }$ we would need more
harmonic amplitudes.

Although the similarity of Eqs. (\ref{rms}) and (\ref{c}) seems to imply
that these two tests are quite equivalent, this is not the case. In the
limiting case $\sigma =0$, the discrete-noise terms give a vanishing
contribution to $C(\gamma ,\sigma )$ because $\sum_\ell a_{N \ell} ^2P_\ell
(\cos \gamma )=0$ for $\gamma \neq 0$, and for the reasons explained in
Fabbri (1992) any dependence on the parameter $bV$ disappears. For finite
values of $\sigma$
this is no more strictly true, but the test based on the shape of $%
C(\gamma ,\sigma )$ remains quite insensitive to $bV$.

A question which is very important for the main issue of our paper is, how
length scales are weighted in Eqs. (\ref{rms}) and (\ref{c}). Since $\Delta
N_{{\rm rms}}^2(\gamma ,\sigma )$ and $C(\gamma ,\sigma )$ depend linearly
on $a_\ell ^2$ , which in its turn is a linear function of $P(k)$, it is
possible to define two window functions $\Psi _F$ and $\Psi _{C\;}$such that
the cosmic-structure contributions to the above quantities are expressed as $%
(2/\pi )\int k^2P(k)\left| \Psi _{F,C}\right| ^2{\rm d}k$. [Cf. Eq. (8) of
Scharf and Lahav (1993).] From Eqs. (\ref{cosmical}), (\ref{rms}) and (\ref
{c}) we get
\begin{eqnarray}
\label{psif}
\left| \Psi _F(k,\gamma ,\sigma )\right| ^2 & = & \frac 1{2\pi
(k^2-4K)^2}  \nonumber \\
  &   & \sum_\ell (2\ell +1)A_\ell ^2F_\ell ^2(\sigma )\left[ 1-P_\ell
(\cos \gamma )\right]
\end{eqnarray}
and
\begin{eqnarray}
\label{psic}
\left| \Psi _C(k,\gamma ,\sigma )\right| ^2 & = & \frac 1{4\pi
(k^2-4K)^2} \nonumber \\
     &  & \sum_\ell (2\ell +1)A_\ell ^2F_\ell ^2(\sigma )P_\ell (\cos
\gamma ).
\end{eqnarray}

Figure 2 reports some results for $\left| \Psi _F\right| ^2$ and $\left| \Psi
_C\right| ^2$ in critical-density models which illustrate general features.
(For consistency with beamwidths used in our tests, we set $\sigma =\gamma $
and $\sigma =2.24^{\circ }$, respectively, for such calculations.) It
appears that small scales are weighted in two of our tests less than in the
work of Scharf and Lahav (1993) using the 2-Jy IRAS survey. This is due to
the statistical bias of the IRAS {\it redshift} survey including 2650
objects. Generally speaking, we expect a large, flux-limited sample to
include more far-away sources than a redshift survey. Both
$\left| \Psi _F\right| ^2$ and $\left| \Psi_C\right| ^2$ increase
monotonically with wavelength. The actual contribution of a
length scale to the signal depends on the spectrum and can be estimated as
$k^3 \left| \Psi _{F,C}\right| ^2 P(k)$. For instance,  for  rms
fluctuations in the angular range $8^\circ - 35^\circ$ the signals
reach their maxima at $\lambda \approx 35 - 50 \, h^{-1}$ Mpc
for a HDM spectrum
with $\lambda _0 = 50h^{-1}$ Mpc, and at
$\lambda \approx 45 - 70 \, h^{-1}$ Mpc
for a CDM spectrum with $\Omega_0h = 0.5$. For the same models, the signals
fall to 10\% of their maxima at $\lambda \approx 7 - 17 \, h^{-1}$ Mpc, i.e.
around the boundary of the linear region, and decrease very rapidly at
smaller scales. Thus nonlinear disturbances
are small.

We further consider the number of local maxima in the density of sources
over the sky. We define $N_{{\rm peak}}(\sigma ,T)$ as the number of peaks
higher than a threshold $T$, counted in the source distribution after
smoothing by a Gaussian beam with dispersion $\sigma $. Following Bond and
Efsthatiou (1987) we take the threshold height $T\;$in units of $\sqrt{%
C(O,\sigma )}$. The peak statistics is described by
\begin{equation}
\label{peak}N_{{\rm peak}}(\sigma ,T)=\frac{f(\gamma _{*}^2,\sigma ,T)}{%
\left( 2\pi \right) ^{\frac 32}\theta _{*}^2},
\end{equation}
\begin{eqnarray}
\label{teta}
\theta _{*}^2 & = & \left[ 2\sum_\ell \ell \left( \ell +1\right)
a_\ell ^2F_\ell ^2\right]  \nonumber \\
      &  &  \left[ \sum_\ell \left( \ell -1\right) \ell \left(
\ell +1\right) \left( \ell +2\right) a_\ell ^2F_\ell ^2\right] ^{-1},
\end{eqnarray}
and%
\begin{eqnarray}
\gamma _*^{2} & = & \left[ \sum_\ell \ell \left( \ell +1\right)
a_\ell ^2F_\ell
^2\right] ^2\left[ \sum_\ell a_\ell ^2F_\ell ^2\right] ^{-1} \nonumber \\
             &   &  \left[
\sum_\ell \left( \ell -1\right) \ell \left( \ell +1\right) \left( \ell
+2\right) a_\ell ^2F_\ell ^2\right] ^{-1},
\label{gamma}
\end{eqnarray}
where the function$\;f$ was calculated by Fabbri (1992) from the theory of
Bond \& Efsthatiou (1987), and $F_\ell \;$now pertains to Gaussian beams.
Inspection of Eqs. (\ref{peak}-\ref{gamma}) shows a highly non-linear
dependence on $a_\ell ^2$ , so that it is not possible to define a window
function. However, the strong weight given to higher order harmonics in Eqs.
(\ref{teta}) and (\ref{gamma}) proves that perturbations with small $\lambda
\;$must have a strong bearing on the statistics for small beamwidths, so
that we expect to be able to provide a lower limit to a dominating scale by
this test. Taking advantage of 300 harmonic amplitudes allowed us to evaluate%
$\;N_{{\rm peak}}$ accurately down to $\sigma \simeq 1^{\circ }$. Since now $%
\sigma \;$denotes a beam dispersion rather than a diameter, the
investigation of small scales is limited somewhat more strongly than for
other quantities.

\section{Data Analysis}

We used a galaxy sample based on the IRAS Point Source Catalog, Version II,
including sources at 60 $\mu $m in the flux interval 0.6-20 Jy. The galaxies
were selected basically using the same criteria as in Strauss et al. (1990),
but we also set the lower flux limit to 0.6 Jy, the completeness limit of
the IRAS survey (Yahil et al. 1986), and extended the region of avoidance
around the galactic plane to $\pm 10^{\circ }$. To reduce the residual
contamination we also excluded other sky regions after source binning; see
below. This sample differs from the one of Fabbri and Natale (1993), not
only for the higher number of sources but also for a different calibration
in Version II of the Catalog, leading to smaller spectral indexes of the
sources. The average spectral index calculated from fluxes at 60 and 25 $\mu
$m is $-1.44$, and the one from fluxes at 60 and 100 $\mu $m is $-1.35$. For
the theoretical modelling in Sect. 2 we adopted the mean value $-1.395$, but
the results are quite insensitive to this choice.

Sources were collected in bins spanning equal intervals $\Delta l$ (i.e., in
Galactic longitude) and $\Delta (\sin b)$ (in the sine of Galactic
latitude). We considered three resolution levels leading to three maps with
128$\times $64, 256$\times $128 and 512$\times $256 bins, respectively, over
the entire celestial sphere. An effective angular diameter $\sigma _b$ such
that $(\pi /4)\sigma _b^2=$ $\Delta l\, \Delta (\sin b)$ can be defined for
the bins. For the above resolution levels we have $\sigma _b=2.24^{\circ }$,
$1.12^{\circ }$ and $0.56^{\circ }$, respectively. For the purpose of our
analysis clean bins with a vanishing number of sources are quite
significant, but we must exclude regions contaminated by galactic emission
or not sufficiently observed during the survey. Hence to each source map we
applied a mask excluding bins which are (i) not observed at least twice in
the survey, or (ii) contained within $\pm 10^{\circ }$ around the galactic
plane, or (iii) confused due to large number of sources (such as the Orion
complex, Taurus and Ophiuchus star forming regions, the Large and Small
Magellanic Clouds), or (iv) contaminated by cirrus emission. Figure 3 gives
the map of our full sample of 12511 sources, obtained by using the above
mask. By means of this procedure, we constructed 3 matrices with indexes ($%
i_b$, $i_l$) spanning Galactic coordinates, whose elements contained either
the number of sources in the bins per unit solid angle $N(i_b,i_l)$ or a
masking flag. These matrices will be referred to as 'bin maps' in the
following.

\begin{figure}
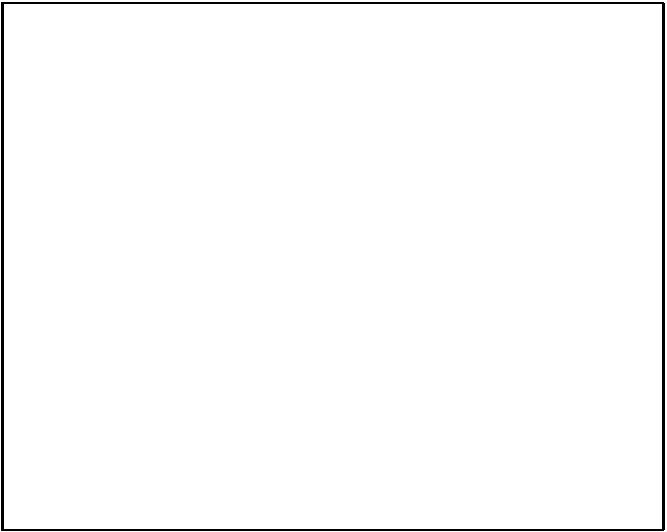

\picplace{7.cm}
\caption[]{Map of the 60 $\mu $m, $0.6-20$ Jy sample}
\label{f3}
\end{figure}
\begin{figure}
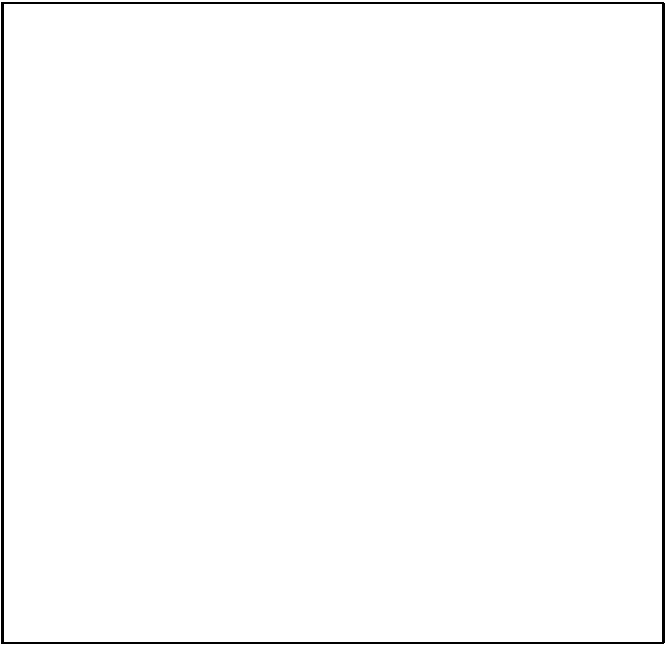

\picplace{8.5cm}
\caption[]{Filled dots give the rms fluctuation
 $\Delta $$N_{{\rm rms}}(\gamma
,\gamma )$ in IRAS PSC for $\sigma _b=2.24^{\circ }$ and $1.12^{\circ }$,
with error bars computed at 1 sigma. The curves represent theoretical models
with parameters selected with best fitting procedures described in Sect. 4.
The full line refers to a hot-matter model with $h\lambda _0=60$ Mpc, $%
\Omega _0=1$, $bV=500$ km/s. The remaining curves describe the following
single-scale models: dashed, $h\lambda _0=40$ Mpc, $\Omega _0=0.7$, $bV=400$
km/s; dotted, $h\lambda _0=40$ Mpc, $\Omega _0=1$, $bV=500$ km/s;
dash-dotted, $h\lambda _0=30$ Mpc, $\Omega _0=0.9$, $bV=400$ km/s}
\label{f4}
\end{figure}
\begin{figure}
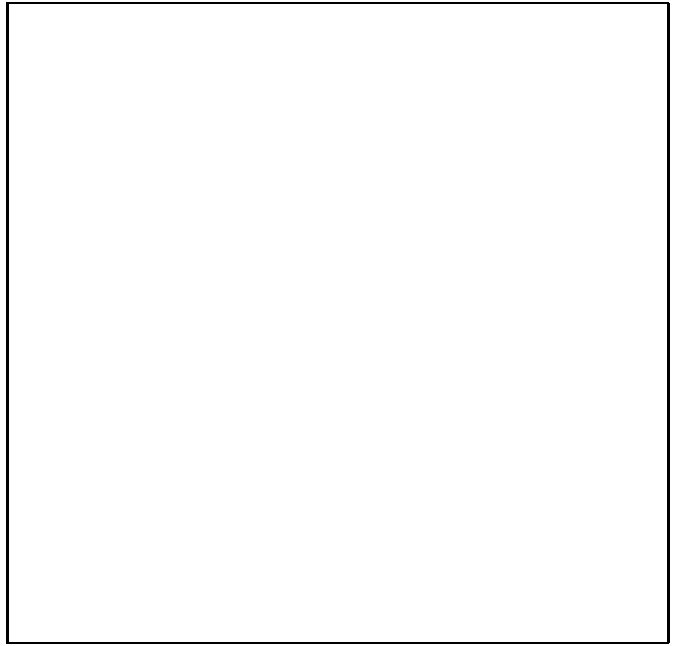

\picplace{8.5cm}
\caption[]{Filled dots give the experimental correlation function $C(\gamma
,2.24^{\circ })$ for $\sigma = \sigma _b=2.24^{\circ }$, with error bars
computed at 1 sigma. The full line describes a hot-matter model with $%
h\lambda _0=45$ Mpc, $\Omega _0=1$, $bV=900$ km/s. Other curves refer to
singlescale models: dashed, $h\lambda _0=30$ Mpc, $\Omega _0=0.9$, $bV=700$
km/s; dotted and dash-dotted, same parameters as in Fig. 4}
\label{f5}
\end{figure}

In the first part of our analysis we compute all the possible differences
between the values of $N(i_b,i_l)$ in adjacent windows and then the rms
value. This amounts to calculating the quantity
$\Delta N_{\rm rms}(\gamma ,\gamma )$
of Eq. (\ref{rms}) for sharp-edge beams, except for the fact that
observation windows are not circular. In order to minimize errors introduced
by non-circular shapes, we maintain beam self-similarity in the generation
of larger beamwidths: Starting from the above bin sizes we merge adjacent
beams so that both $\Delta l$ and $\Delta (\sin b)$ are doubled at each
step. The effective beam diameter $\sigma \;$covers the range from $\sigma _b
$ to $71.7^{\circ }$ at logarithmically spaced intervals. This procedure is
similar to the one described in Fabbri \& Natale (1990). The present
analysis only differs in the treatment of masked bins in the merging
process: Composed windows are now rejected when more than one bin is masked,
otherwise the masked bin is filled with the average number of sources
evaluated on all unmasked bins in the whole map. The results obtained for $%
\sigma _b=1.12^{\circ }$ and $2.24^{\circ }$ are given in Fig. 4 with error
bars at 1 sigma. Ideally the two data sets should coincide up to a
multiplicative factor. The figure shows that this is verified to a good
accuracy: Quite small deviations arise from differences in the masks
constructed at different resolutions. The data set with $\sigma
_b=0.56^{\circ }$ is not reported here, since it gives no new results and
the first data point could not be compared to theory for the reasons
explained in Sect.2.

\begin{figure}
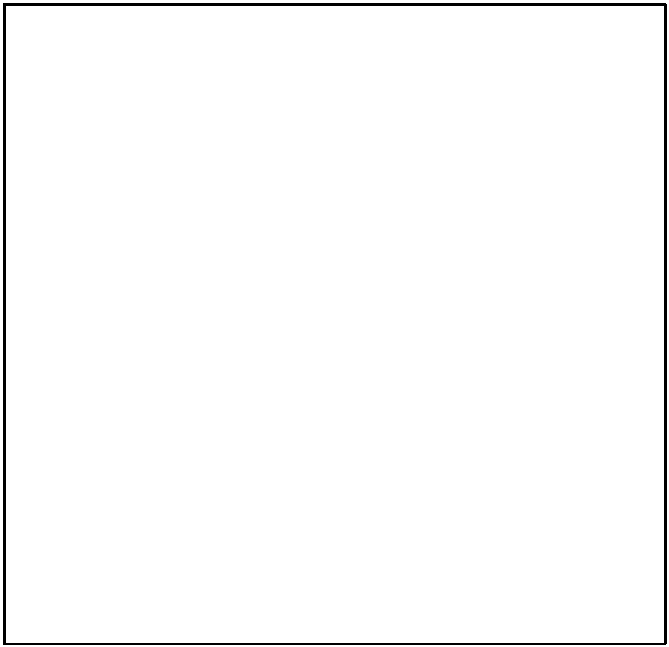

\picplace{8.5cm}
\caption[]{The peak number vs. threshold $T$ and Gaussian beam dispersion $%
\sigma \;$in IRAS PSC. Data points, normalized to the entire sky, are
represented by mesh nodes. Three distinct meshes are given, corresponding to
$\sigma _b/\sigma = 0.93$ (a), 0.47 (b), 0.23 (c)}
\label{ex6a}
\end{figure}
\begin{figure}
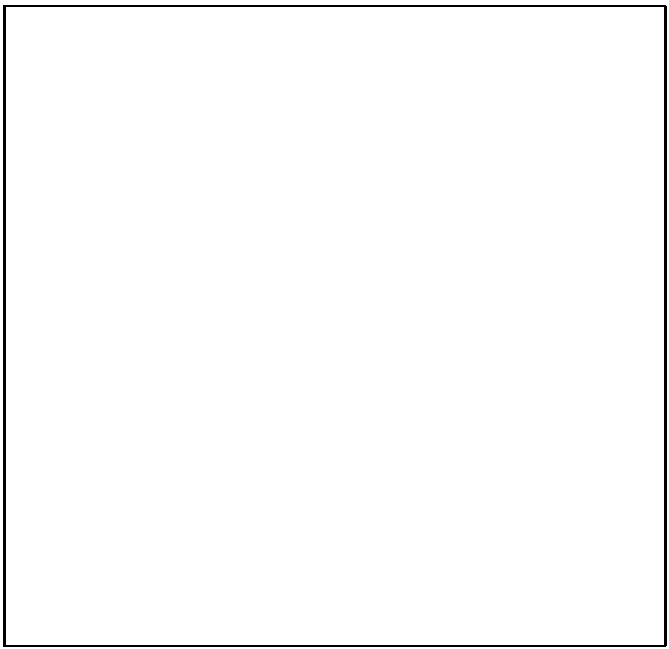

\picplace{8.5cm}
\caption[]{The peak number for $\sigma _b/\sigma =0.47$ from the North and
South halves of the sample, labelled by N and S respectively. Data points,
normalized to one half of the sky, are represented by mesh nodes}
\label{ex6b}
\end{figure}

\begin{figure}
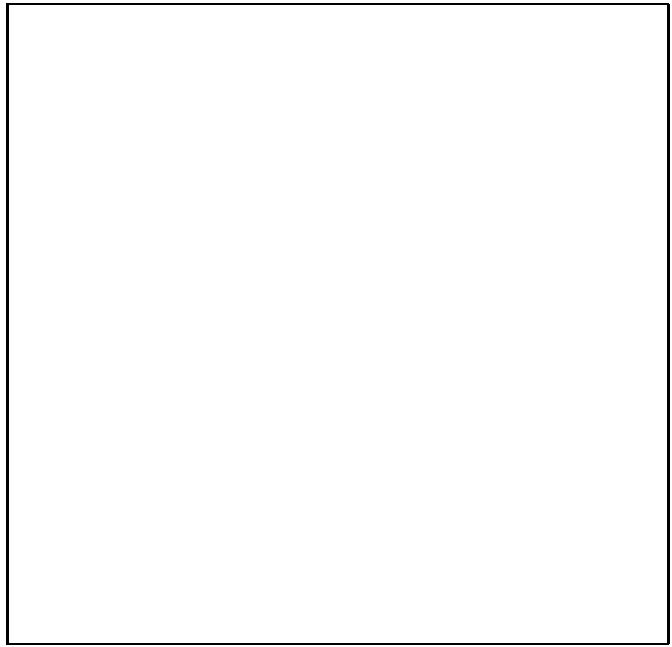

\picplace{8.5cm}
\caption[]{The peak number vs. threshold $T$ for $\sigma _b/\sigma =0.47$.
Filled dots represent experimental data extrapolated to the entire sky, with
1-sigma error bars, at three beamwidths $\sigma =1.2^{\circ }$, 2.4$^{\circ
}$ and 4.8$^{\circ }$ (with $N_{{\rm peak}}$ decreasing for increasing $%
\sigma $). Theoretical curves are given for single-scale models with the
following parameters: full line, $h\lambda =20$ Mpc, $\Omega _0=0.9$, $bV=500
$ km/s (best fit for this set of data); dashed, $h\lambda =40$ Mpc, $\Omega
_0=1$, $bV=500$ km/s; dotted, $h\lambda =30$ Mpc, $\Omega _0=0.9$, $bV=400$
km/s}
\label{ex7a}
\end{figure}
\begin{figure}
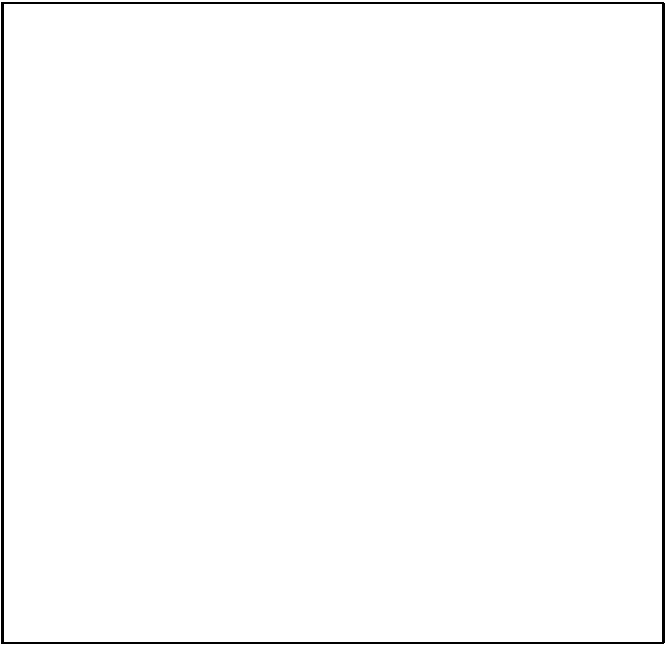

\picplace{8.5cm}
\caption[]{The peak number vs. threshold $T$ for $\sigma _b/\sigma =0.23$.
Filled dots represent experimental data, with 1-sigma error bars, at
beamwidths $\sigma =2.4^{\circ }$, 4.8$^{\circ }$ and 9.6$^{\circ }$ (with $%
N_{{\rm peak}}$ decreasing for increasing $\sigma $). The full line
describes the single-scale model with $h\lambda =50$ Mpc, $\Omega _0=0.9$, $%
bV=800$ km/s (best fit for this set of data); the dashed and dotted lines
refer to the same models as in the preceding figure}
\label{ex7b}
\end{figure}

We also computed the correlation function at $\sigma =\sigma _b=2.24^{\circ }
$ by means of the equation
\begin{eqnarray}
\label{cexp}
C(\gamma ,\sigma _b) & = & \left[ n_p(\gamma ,\sigma _b)\right]
^{-1}  \nonumber \\
 & & \sum \left[ N(i_b,i_l)-\left\langle N\right\rangle \right] \left[
N(j_b,j_l)-\left\langle N\right\rangle \right] ,
\end{eqnarray}
where the summation is performed over all pairs of beams such that the
angular separation of center coordinates lies in the range $(\gamma -\frac
12\sigma _b,\;\gamma +\frac 12\sigma _b)$. The number of such pairs is $%
n_p(\gamma ,\sigma _b)$, and $\gamma =n\sigma _b$ with $n$ an integer.
Masked bins are simply skipped in this computation. Now we cannot maintain
self-similarity; therefore we excluded two Galactic polar cups at $\left|
b\right| >60^{\circ }$ where the stretching of bin shapes is strong. Figure
5 reports data in the range $\gamma \in (2.24^{\circ },24.7^{\circ })$.

In order to treat the peak number statistics, we should take into account
that theoretical predictions are available for smooth distributions over the
celestial sphere. For any distribution of discrete sources we encounter the
following problem, that moving a sharp-edge beam across the sky the detected
source number varies in a discontinuous manner. Therefore we must limit
ourselves to smooth beams, and in this paper we shall consider Gaussian
beams. In principle, for each beam position in the sky we should weight all
sources in the sample by a smoothing Gaussian factor $\exp \left[ -\frac
12\left( \vartheta /\sigma \right) ^2\right] $, with $\vartheta \;$the
source angular separation from the beam center. In practice it is more
efficient to replace the actual content of each bin by an effective, smoothed
number of sources according to the formula
\begin{equation}
\label{peakexp}
N_{{\rm sm}}\left( i_{b0},i_{l0}\right) =\frac{%
{\sum } N\left( i_b,i_l\right) \exp \left[ -\frac
12\left( \vartheta _{i_b i_l} /\sigma \right)
^2\right] }{{\sum } \exp \left[ -\frac 12\left( \vartheta
_{i_b i_l} /\sigma \right) ^2\right] },
\end{equation}
where $\vartheta _{i_b i_l} =
\gamma \left( i_b,i_l,i_{b0},i_{l0}\right) $ is the angular
separation between cells $(i_b,i_l)$ and $(i_{b0},i_{l0})$. Masked bins
obviously are disregarded. Now a complication arises from the fact that the
measured number of peaks explicitly depends on both $\sigma \;$and $\sigma _b
$, while theory assuming a continuous sampling of the source number-density
field only predicts the $\sigma $-dependence. Therefore, to consistently
compare data at varying values of $\sigma \;$we had to require similar
geometries, i.e. equal ratios $\sigma _b/\sigma $. Since we considered four
logarithmically spaced values of $\sigma \;$in the range $(1.2^{\circ
},9.6^{\circ })$, from our bin maps we can build three sets of data with $%
\sigma _b/\sigma =0.23$, 0.47, and 0.93 respectively. Each set includes a
triplet of values of $\sigma $; for instance, for $\sigma _b/\sigma =0.23$
we can collect data at $2.4^{\circ }$, $4.8^{\circ }$ and $9.6^{\circ }$
taking advantage of all of our bin maps.

In practice starting from 3 original bin maps we must build 9 smoothed maps,
corresponding to specific values of $\sigma \;$and $\sigma _b$, and in each
of them we search for local maxima of the $N_{{\rm sm}}$ distribution.
The search was performed comparing the source number of each bin with those
of eight contiguous bins. We handled with the presence of holes in the map
developing algorithms which extended the analysis to larger neighborhoods of
potential maxima. The presence of very few uncertain cases does not affect
our results. The identified peaks were then normalized to the rms
source-number fluctuation calculated for each smoothed map, and the
distribution $N_{{\rm peak}}(\sigma ,T)$ was computed for 7 equally spaced
values of $T$ between 0.1 and 1.6. Since the number of peaks should then be
extrapolated to the whole sky, we have to evaluate the effective area
associated to the distribution of detected peaks. We adopt two procedures:
(a) We only include sky bins connected to the two big unmasked continents at
$\left| b\right| >10^{\circ }$ and surrounded by no more than 5 masked bins,
and (b) we accept all unmasked bins.

Figure \ref{ex6a} gives $N_{{\rm peak}}(\sigma ,T)$, extrapolated to the
whole sky from the entire set of unmasked bins using the first method above.
Since only data sets with fixed $\sigma _b/\sigma $ are homogeneous, the
figure shows 3 distinct meshes. For the case $\sigma _b/\sigma =0.93$ only
two values of $\sigma$
are considered because data at $\sigma =0.6^{\circ }$ could
not be compared with theory. Accepting all unmasked bins for the computation
of areas slightly lowers the peak numbers, but this has no important effect
as will be seen in Sect. 4. Figure \ref{ex6b} reports $N_{{\rm peak}}(\sigma,
T)$ for the case $\sigma _b/\sigma =0.47$, derived from the North and
South halves of the sample. In this figure the peak number is extrapolated
to one half of the sky. Error bars at 1 sigma are evaluated as $\Delta N_{%
{\rm peak}}=\sqrt{N_{{\rm peak}}/f_{{\rm u}}}$, with $f_{{\rm u}}$ the
unmasked sky fraction. Such errors are reported in Figs. \ref{ex7a} and \ref
{ex7b}.

\section{Results of Best Fits}

For each of our tests we basically followed the same procedure which can be
outlined as follows. After selecting values for the parameters describing
the source properties, we are left with two free parameters related to
cosmic inhomogeneities, i.e. the products $h\lambda $ (or $h\lambda _0$) and
$bV$. For single-scale models one more free parameter is the cosmic density
parameter $\Omega _0$; for HDM models we set $\Omega _0=1$, and for CDM
models the spectral properties only depend on the shape parameter $\Omega _0h
$. For each parameter set we first minimize $\chi ^2$ with respect to a
normalization factor, which in the $\Delta N_{{\rm rms}}^2\;$and $C(\gamma
,\sigma )$ tests determines the normalization of $P(k)$. For the peak
statistics this factor should not be fitted in the ideal case of a pointlike
sampling of a continuous field, namely for $\sigma _b/\sigma \rightarrow 0$,
but in our case $\sigma _b/\sigma $ can be as large as 0.93, and we have no
warrant that theoretical predictions should reproduce finite-bin data
exactly. However the normalization factor turned out to be usually close to
unity (within $\sim $$10\%),$ and it was always so in the neighborhoods of
the absolute minima. We then obtain 2- or 3-dimensional $\chi ^2\;$grids, in
which we search for $\chi _{\min }^2$. Statistical errors on each parameter
are calculated projecting out the extremal points of ($\chi _{\min }^2+1$)
contours on the coordinate axes of two-parameter planes, after one more
minimization with respect to the third, unplotted parameter when available.
The errors quoted in the following tables are quadratic combinations of such
1-sigma errors and grid half spacing.

\subsection{Single-scale models}

For single-scale models, adopting the optimal source parameters we
calculated a 3-dimensional lattice of equally spaced points in the ranges$%
\;h\lambda =10-100$ Mpc, $\Omega _0=0.1-1$ and $bV=100-900$ km/s. Table 1
lists fitted parameter values that we obtained from several tests, as well
as $\chi _{{\rm min}}^2$ compared to the number of degrees of freedom. Since
errors on $\Omega _0$ often exceeded the range investigated, we enclosed the
best values within parentheses to declare their loss of significance.

\renewcommand{\arraystretch}{1.5}
\begin{table*}
\caption[]{Best fit parameters for single-scale models}
\begin{flushleft}
\begin{tabular}{lllll}
\noalign{\smallskip}
\hline \noalign{\smallskip}

Case$\,^{\dagger }$ & $h\lambda $ (Mpc) & $\Omega _0$
& $bV$ (km/s) & $\chi _{{\rm min}}^2/$DOF     \\
\noalign{\smallskip}
\hline \noalign{\smallskip}

F$1^{\circ }-36^{\circ }(1^{\circ })$ & 40$_{-12}^{+8}$ &(1.0)&
 500$_{-360}^{+50}$ &4.1/2 \\

F$1^{\circ }-72^{\circ }(1^{\circ })$ & 40$_{-7}^{+15}$ & $ 0.7\pm 0.5 $
&400$_{-130}^{+180}$
&15.2/3  \\

F$2^{\circ }-72^{\circ }(2^{\circ })$ & 40$_{-8}^{+16}$  & $0.7 \pm 0.5  $
&400$_{-130}^{+150}$
&14.5/2 \\

C$2^{\circ }-25^{\circ }$ &30$_{-6}^{+12}$ & (0.9)
& 700$_{-600}^{+240}$  & 21.1/7  \\

P$2.4^{\circ }$(a) & 20$_{-5}^{+13}$ &  (0.9) &500$\pm 410$ &14.2/17 \\

P$2.4^{\circ }$(a, N)$\,^{\S}$ & 30$_{-9}^{+21}$ & (1.0)&
 300$_{-240}^{+280}$  &15.6/17\\

P$2.4^{\circ }$(a, S)$\,^{\S}$ &20$_{-5}^{+15}$ & (0.8)
& $600 \pm 350 $ &20.8/17 \\

P$2.4^{\circ }$(b) &20$_{-5}^{+12}$ & (0.9)& 500$_{-410}^{+400}$
& 14.6/17 \\

P$4.8^{\circ }$(a) &50$_{-8}^{+17}$ &(0.9)& 800$_{-530}^{+130}$
&24.0/17 \\

P$4.8^{\circ }$(b) &50$_{-8}^{+16}$ & (0.9) &800 $_{-530}^{+130}$
 &24.5/17 \\

P$1^{\circ }-10^{\circ }$(a)$\, ^{\ddagger }$ & 30$_{-7}^{+10}$ &(0.8)
 & 400$_{-300}^{+240}$
 &30.0/22 \\

F+P$\, ^{*}$ &40$_{-10}^{+9}$ & $1.0 \pm 0.6$ & 500$_{-230}^{+50}$
& 47.3/27 \\

C+P$\, ^{*}$ &30$_{-6}^{+10}$ &(0.9) &400$_{-290}^{+310}$
 & 51.2/32 \\
\noalign{\smallskip}
\hline
\end{tabular}

$^{\dagger }\;$Label F denotes rms fluctuation tests in the specified
beamthrow range;
rounded-off bin size is within parentheses. C means correlation test in the
specified beamthrow range; bin size is $2.24^\circ $. P marks
 peak number tests
for triplets of $\sigma$ around quoted values. For the meaning of (a) and (b)
labels
see Sect. 3.

$^{\S}\;$N , S = Northern and Southern half of the sky.

$^{\ddagger }\;$Full range of $\sigma $
used.

$^{* }\;$Joint tests combine P$ 1^\circ - 10^\circ $(a) data with
F$ 1^\circ - 72^\circ (1^\circ )$ and C
$2^\circ - 25^\circ$, respectively.
\end{flushleft}
\end{table*}

Results reported in the table for the rms fluctuation tests (denoted by the
letter F) refer to two beamthrow ranges, 1.12$^{\circ }-35.8^{\circ }$ and
2.24$^{\circ }-71.7^{\circ }$, and two values of $\sigma _b$. It was
especially important to check a possible dependence of the best-fitted
length scale on the beamthrow range. No difference appears in the quoted
best values of $h\lambda $, but $\chi _{\min }^2\;$is high in the upper
range for the difficulty of fitting the data point at $71.7^{\circ }$. All
of the rms fluctuation and correlation function data (see the fourth row in
the table) are consistent with $h\lambda \simeq 30-40$ Mpc and $bV\simeq
300-500$ km/s. The parameter $bV$ is correlated to $\Omega _0$; if we assume
$\Omega _0=1$ (which is not required by our data) we can also derive a
preferred range $bV\simeq 400-500$ km/s. Results from the peak statistics
reported at rows 5-10 of the table (denoted by the letter P) are associated
to a beamwidth value. Each entry is in fact obtained using 3 beamwidths (of
which the label value is the geometric mean) referring to data points with a
single value of $\sigma _b/\sigma $. Thus for instance, the cases labelled
by P$2.4^{\circ }$(a) and P$2.4^{\circ }$(b) use each 21 data points with $%
\sigma = 1.2^{\circ }$, 2.4$^{\circ }$ and 4.8$^{\circ }$ and $\sigma
_b/\sigma =0.47$. The letters a and b refer to the methods for extrapolating
the number of peaks to the whole sky described in Sect. 3. The test
combining all of our $N_{{\rm peak}}$ data is denoted by the full range $%
1^{\circ }-10^{\circ }$. Entries in the 6th and 7th row show that consistent
results were obtained for the North or South hemispheres (which was verified
for $N_{{\rm peak}}$ tests at all beamwidths). We do not report results for
the self-similar case with $\sigma _b/\sigma =0.93$ because no significant
result could be found after exclusion of data with $\sigma = 0.6^{\circ }$.
Inspection of the table shows that in all cases we consistently find $%
h\lambda =20-30$ Mpc and 50 Mpc for smaller and larger beamwidths,
respectively. The former can be regarded as a safe lower limit to a
preferred scale.

\begin{figure}
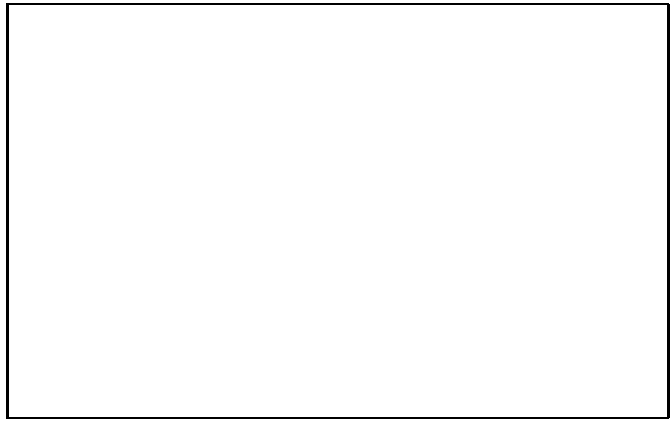

\picplace{5.5cm}
\caption[]{The $\chi ^2\;$contours in the $(h\lambda ,\;bV)$ plane
corresponding to levels 2.30 (full lines) and 6.17 (dashed) above $\chi
_{\min }^2$, for a few tests based on single-scale models. The specific
tests involved are identified by the curve labels, which are also used in
Table 1 and explained in the text}
\label{ex8}
\end{figure}

Figure \ref{ex8} gives a few contours in the $(h\lambda ,\;bV)$ plane
corresponding to $\chi _{\min }^2+\Delta \chi ^2$, with $\Delta $$\chi ^2=2.3
$ and 6.17, which for a Gaussian field should delimit the ellipses at
confidence levels of 1 and 2 sigmas, respectively [cf. Press et al. (1988),
Chapt. 14]. The figure shows that contours for the P tests do not overlap at
2 sigmas, since the preferred length scale show some dependence on angular
scale. However, all of the other tests (included those not reported in the
Figure) provide intermediate results.

Finally, since as already remarked $N_{{\rm peak}}$ weights high-order
harmonics much more strongly than $\Delta N_{{\rm rms}}(\gamma ,\gamma )$
and $C(\gamma ,\sigma )$ do, we treated $N_{{\rm peak}}$ data as quite
independent from the other two sets of data. This allows us combining
couples of tests, simply summing up the respective $\chi ^2$ fields. Entries
labeled as F+P and C+P in the last two rows of Table 1 respectively combine
data from the P$1^{\circ }-10^{\circ }$(a) case with data from F$1^{\circ
}-72^{\circ }(1^{\circ })$ and C$2^{\circ }-25^{\circ }$. The resulting $%
\chi _{{\rm min}}^2$ are only slightly higher than the sums of minima in the
component fields, which obviously refer to different values of the fitted
parameters: This fact is due to the slow variations of $\chi ^2$ around the
minima. Therefore the rather large values of $\chi _{{\rm min}}^2$ in Table
1, which emphasize the limits of the single-scale approximation, mainly
arise from some difficulties that the models encounter in each of our basic
tests.

\subsection{Continuous spectra}

In the case of continuous-spectrum models after minimization of $\chi ^2$
with respect to the normalization factor we built 2-dimensional grids in the
plane $(h\lambda _0,\, bV)$ or $(\Omega _0h,\, bV)$
for HDM and CDM respectively.

\begin{table*}
\caption[]{Best fit parameters for HDM models}
\begin{flushleft}
\begin{tabular}{llllll}
\noalign{\smallskip}
\hline \noalign{\smallskip}

Case & $h\lambda _0$ (Mpc) & $bV$ (km/s) & $b\sigma _{8V}\,^{\dagger}$
   & $b\sigma
_{8N}\,^{\ddagger }$ & $\chi _{{\rm min}}^2$/DOF    \\
\noalign{\smallskip}
\hline \noalign{\smallskip}

 F$1^{\circ }-36^{\circ }(1^{\circ })$ & $60_{-14}^{+32}$
 &  $500_{-70}^{+130}$
 &  0.60 & 0.75 & 2.5/3  \\

C$2^{\circ }-25^{\circ }$ &  $45_{-4}^{+11}$ & $900\pm 560$
 & 1.14 & 1.23 & 8.3/8    \\

P$1^{\circ }-10^{\circ }$(a) &  $42.5_{-4.5}^{+17}$ & $700_{-190}^{+130}$
 & 0.89 & --
 & 25.6/23 \\

F+P$\,^{*}$ & $52.5_{-7}^{+22}$ & $500_{-60}^{+90}$ & 0.62
 & -- & 30.3/27               \\

C+P$\,^{*}$ & $45.0_{-1.5}^{+5}$ & $700_{-270}^{+130}$ & 0.89
 & -- &34.1/33            \\
\noalign{\smallskip}
\hline
\end{tabular}

$^{\dagger }\;$Calculated from $bV\;$in the preceding column.

$^{\ddagger }\;$Calculated from the normalization factor of $P(k)$.

$^{*}\;$Joint tests combine P$1^{\circ }-10^{\circ }$(a) data with F$%
1^{\circ }-36^{\circ }(1^{\circ })$ and C$2^{\circ }-25^{\circ }$,
respectively.
\end{flushleft}
\end{table*}

\begin{table*}
\caption[]{Best fit parameters for CDM models$\,^{\dagger}$}
\begin{flushleft}
\begin{tabular}{lllll}
\noalign{\smallskip}
\hline \noalign{\smallskip}

Case  & $\Omega_0h$ & $b V$ (km/s) & $\chi^2_{\rm min}$/DOF
  & $\chi^{*2}_{\rm min}$/DOF \\
 \noalign{\smallskip} \hline  \noalign{\smallskip}

F$ 1^\circ - 36^\circ (1^\circ )$ & $ 0.10^{+ 0.03}_{- 0.10}$
 & $ 600^{+ 150}_ {-90}$ & $ 5.7 / 3 $ & $21.1/ 3$ \\

C$ 2^\circ - 25^\circ $ & $ 0.50^{+ 0.10}_{- 0.15} $
& $ 900^{+ 50}_{- 660} $ & $ 13.8 / 8 $ & $13.8 /8$ \\

P$ 1^\circ - 10^\circ (a) $ & $ 0.35^{+ 0.18}_{- 0.15}$
& $ 900^{+ 150}_{- 160} $ & $ 23.5 / 23 $ & $24.9 /23$ \\

F+P$\, ^* $ & $ 0.15^{+ 0.03}_{- 0.15}$ &
$  600^{+ 160}_{- 60}$ & $  35.2 / 27 $ & $45.6/27 $ \\

C+P$\, ^* $ & $ 0.45^{+ 0.12}_{- 0.13}$
& $ 900^{+ 130}_{- 240} $ & $ 38.2 / 33 $ & $ 38.8 /33  $ \\
\noalign{\smallskip}
\hline
\end{tabular}

$ ^\dagger \;$Starred quantities (last column) refer to best fits
 limited to the range $\Omega_0h \ge 0.5 $.

$ ^* \;$Joint tests combine P$1^\circ - 10^\circ$(a) data with
  F$1^\circ - 36^\circ (1^\circ )$ and C$2^\circ -
 25^\circ $, respectively.
\end{flushleft}
\end{table*}

For HDM models the range limits were the same as for single-scale models,
but the 2-dimensional $\chi ^2$ grid was not equally spaced in the $h\lambda
_0$ direction. (The finest spacing, adopted around the $\chi ^2$ minima, was
2.5 Mpc.) Table 2 reports the results obtained in a subset of our best
fitting procedures, which is sufficient to illustrate our findings. The
table shows that the values of $\chi ^2$ are very close to the numbers of
degrees of freedom for all of the reported tests. The best fitted values of $%
b\sigma _8$, derived both from $bV\;$and from the normalization factor of $%
P(k)$, are often fairly close to the 'canonical' value 0.7 (Lahav et al.
1990, Saunders et al. 1992) although there is some spreading in the results.
One might have expected to find lower values of $b\sigma _{8\;}$for HDM
models; however, this is not the case because our spectra are peaked at
low wavelengths. The only real limitation for so satisfactory results is the
exclusion of the data point at $71.7^{\circ }$ in the case of the rms
fluctuation test. The behaviour of the HDM model represented by the full
line in Fig. 4 shows how such a point spoils the quality of a fit which
would otherwise be very good. In general, the quality of fits for any
sharply-peaked (including single-scale and double-power) models is somewhat
worse at larger angles. Thus in Fig. 5 models do not strictly follow the
oscillations of $C(\gamma ,2.24^{\circ })$ around $\gamma \approx %
20^{\circ }$, and finally, all of our theoretical distributions of density
peaks slightly deviate from experiment at our largest beamwidth, the
measured $N_{{\rm peak}}(9.6^{\circ },T)$ being systematically lower than
predicted for $T>0.8$. We could not extend the analysis to angles larger
than $10^{\circ }$ because the small number of expected peaks is completely
masked by Poisson noise.

\begin{figure}
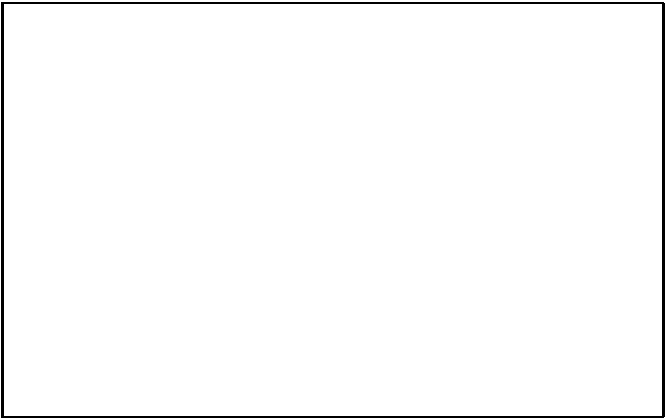

\picplace{5.5cm}
\caption[]{The $\chi ^2\;$contours in the ($bV,\, h\lambda _0$) plane
corresponding to levels 2.30 and 6.17 above $\chi _{\min }^2$, for two
global tests on HDM models, respectively labelled by F+P and C+P}
\label{ex9}
\end{figure}

For the above reasons, we performed the global test F+P for HDM combining
the P$1^{\circ }-10^{\circ }$(a) and F$1^{\circ }-36^{\circ }(1^{\circ })$
data. Contours at 1- and 2-sigma confidence levels for the F+P and C+P tests
are reported in Fig. \ref{ex9}. These global tests appear to be somewhat
complementary, setting more stringent limits on $bV$ and $h\lambda _0$
respectively. We can satisfy both simultaneously with $bV\simeq 450-550$
km/s and $h\lambda _0\simeq 45-55$ Mpc.

\begin{figure}
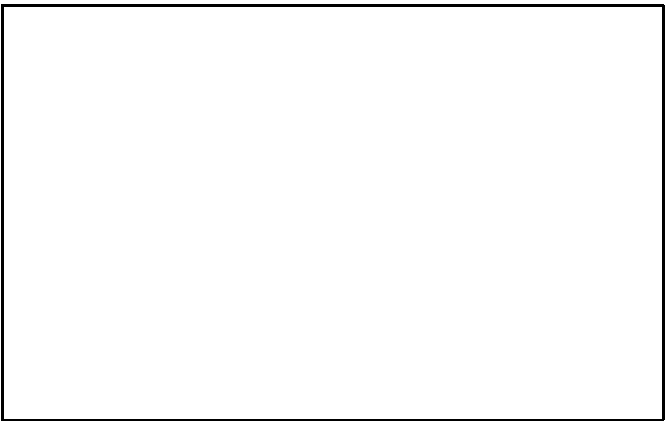

\picplace{5.5cm}
\caption[]{The $\chi ^2$contours in the $(bV,\, \Omega _0h)$ plane
corresponding to the same levels above $\chi _{\min }^2\;$as in the
preceding figure, for two global tests on CDM models, respectively labelled
by F+P and C+P}
\label{ex10}
\end{figure}

Table 3 collects results for CDM tests corresponding to those of Table 2. In
all of these tests $\Omega _0h$ is allowed to vary in the range (0.1, 1.2).
However in the last column we also report $\chi _{{\rm min}}^{*2}$ , the
minimum chi-square obtained in the smaller range (0.5, 1.2). The latter
range is suitable to test critical density models, letting $h\;$vary; in
this case the constrained minima are always found at $h=0.5$, but the fits
are unsatisfactory with the exception of peak statistics. Allowing for
low-density models the situation improves substantially: For tests involving
peak statistics only, $\chi _{{\rm min}}^2$ is even slightly smaller than
for hot matter. However, the chi-square minima identified by our tests are
spread in the range $\Omega _0h=0.1-0.5$ and do not appear so mutually
consistent as in the case of hot matter. The situation is made clear by
inspection of Fig. \ref{ex10}: For cold matter the C+P and F+P tests allow
us identifying an overlap region ($\Omega _0h\simeq 0.4,\;bV\simeq 600$
km/s) only at a formal confidence level of 2 sigmas.

In order to decide whether our tests really argue against CDM models in
low-density cosmologies, we should discuss the significance of error bars we
attach to fitted parameters. The confidence regions derived from
$\chi^2$ contours
stem from purely statistical errors. (Error bars quoted in Tables 1-3 also
include grid spacing.) As pointed out by White, Krauss and Silk (1993)
cosmic variance can be more important than instrumental noise in anisotropy
experiments such as those on the cosmic background radiation. In our case
the 'instrumental' noise (i.e., the error in counting extragalactic sources
in sky bins in a definite flux interval) is negligible in front of the noise
term $\Delta $$N(i_b,i_l)=\sqrt{N(i_b,i_l)}$, which is the source of our
error bars. Cosmic variance is just this noise term. In order to check that
this interpretation is correct, we attributed the standard cosmic-variance
uncertainty $\Delta $$(a_l^2)=a_l^2/\sqrt{(l+\frac 12)}$ to all of the
models which best fitted any one of our tests, and derived the corresponding
uncertainties $\Delta _{CV}$ on the theoretical quantities $\Delta $$N_{{\rm %
rms}}^2$, $C(\gamma ,\sigma )$ and $N_{{\rm peak}}$. We found that our
declared errors in Figs. 4-\ref{ex7b} are typically $0.7-1.7$ times $\Delta
_{CV}$, and this ratio is larger as it must when theoretical predictions lie
below experimental data points. The average ratio $\sim 1.2$ $\sim f_{{\rm %
u}}^{-\frac 12}\;$is consistent with theoretical expectation taking into
account the unmasked portion of the sky.

As noted by Scaramella and Vittorio (1993), taking cosmic variance into
account in the computation of the $\chi ^2$ field does not provide exactly
the same results as simulations: Models predicting anisotropies larger than
experiment are affected by a larger cosmic variance, and can be rejected
less easily than we could infer from inspection of the $\chi ^2$ field. From
these considerations and the inspection of Table 3 and Fig. \ref{ex10}, it
follows that low-density CDM models are not clearly ruled out by our tests.
However, very simple HDM models work remarkably better.

\section{Discussion}

The results expounded in Sect. 4 show that a common picture arises from both
single-scale and HDM models. The scale $\lambda \sim 30-40\;h^{-1}$ Mpc
suggested by the former is only slightly smaller than the peak wavelength $%
\lambda _0$ arising from the latter (in particular, the value $45-55\;h^{-1}$
Mpc implied by the simultaneous consideration of the global tests). The
linear component of our local velocity (in practice, the velocity of the
Local Group) is estimated around 500 $b^{-1}$ km/s, in agreement with
current views on the large-scale velocity field, although not with inferences
of about 1000 km/s (cf. Collins et al. 1986). Our basic conclusions remain
unchanged for different source parameters and sharply-peaked spectral shapes
provided only that IRAS sources already existed at $z\sim 0.1$. The quality
of HDM fits suggests that $P(k)$ is characterized by some length scale
smaller than appearing from the analyses of redshift surveys.

Possible objections to this result might involve the reliability of the
sample, the ability of our tests to effectively probe the full perturbation
spectrum and contamination from small, non-linear scales.

As to the first point, we are confident that our sample is reliable. Similar
results (using single-scale models and two of the present tests) were
obtained from another sample derived from IRAS PSC, Version I with more
conservative selection criteria (Fabbri \& Natale 1990, 1993). Thus there is
no chance that our results might be affected by star contamination. No
problem should arise from the finite depth of the sample. Applying the peak
statistics separately to the South and North halves of the sky we reduced
the effective linear scale of the sample by a factor 2$^{\frac 13}$ , but we
observed no shift of the best-fitted wavelength or the confidence regions to
lower scales. Also, our models carefully take into account the flux interval
of the sample, the source luminosity function and the relativistic
expression of the luminosity distance; they explicitly allow for the
effective depth arising from the interplay of such factors. So there is no
reason why such a length scale should strongly affect our tests, nor why a
spectrum like (\ref{hot}) should fit data better than equation (\ref{cold}).
[In fact, the luminosity function of Saunders et al. (1990) does not even
allow us identifying a length scale so sharply.] The depth of the sample
could drastically affect the fittings only if it arised from a strong
cutoff, totally independent of the above factors and effective at redshifts $%
<$ 0.1. There is no evidence for such a cutoff. We previously studied the
effect of varying the lower flux limit on Version I of PSC and found no
effects of this kind (Fabbri \& Natale 1990). We further notice that Jing \&
Valdarnini (1993) use the redshift field of 2-Jy IRAS galaxies to infer a
peak wavelength of $\sim 300h^{-1}$ Mpc. The median depth of their sample
is quite smaller than ours. Thus the different results arise from the
different techniques, not from peculiarities of our sample.

It is therefore important to notice that the shape of the window functions
which can be defined for our tests do not weight small length scales to a
pathological extent; actually they weight them less than a 2-dimensional
analysis based on IRAS redshift survey (Scharf and Lahav 1993).

All of the above considerations cannot prove of course that our best value
for $\lambda _0$ should be the location of the absolute maximum of $P(k)$.
We are aware that best-fit procedures are not equivalent to directly
constructing the shape of $P(k)$ from data. This consideration is especially
important because we find that our tests, when applied to CDM models, point
to larger length scales. This apparently strange result may also be linked
to the fact that in 2-dimensional analyses length scales are smeared out;
under this respect 3-dimensional studies (e.g. Feldman et al. 1994) are
superior. However, the quality of our HDM low-wavelength fits and the
arguments discussed in the Introduction makes us suspect that current
spectra obtained from redshift catalogs may not represent the correct $P(k)$
in real space very accurately. Although there is a qualitative agreement
between spectra inferred from visible and IRAS galaxies and clusters, a
close inspection of the papers by Fisher et al. (1993), Einasto et al.
(1993) and Jing \& Valdarnini (1993) shows significant differences between
the spectra proposed by these authors. A conservative interpretation of our
results is that at least a spectral feature should exist, not detected in
the above works. Strictly speaking, a local peak around $45h^{-1}$ Mpc
appears in Fig. 5 of Jing \& Valdarnini, but we believe its strength is too
weak to justify our findings. A recent work by Baugh \& Efstathiou (1993)
studies the 2-dimensional correlation function in the APM survey and
recovers $P(k)$ by the method of Lucy (1974). Their spectrum rises above a
power law around 30$h^{-1}$ Mpc and appears to be peaked at 100$h^{-1}$ Mpc.
Although the authors find that their results may be compatible with previous
work, still they declare that the discrepancy showed by their Fig. 11 is
disturbing.

In our opinion the support for a steady increase of $P(k)$ up to very large
wavelengths is not overwhelming. [Difficulties may arise also in connection
with data on the anisotropies of the cosmic background radiation (Fabbri \&
Torres 1994)]. The problems arising from the comparison of results found
with different techniques may indicate that the cosmic spectrum is more
complicated than the smooth theoretical shapes appearing in the current
literature.

\begin{acknowledgements}We wish to thank M. T. Dibari for useful discussions on
IRAS PSC. This work is partially supported by Agenzia Spaziale Italiana
under Contract \# 92RS64, and by the Italian Ministry for the University and
Scientific and Technological Research (Progetti Nazionali e di Rilevante
Interesse per la Scienza).\end{acknowledgements}


\begin{thebibliography}{99}
\bibitem{}  Bahcall, N.A., Cen, R., Gramann, M. 1993, ApJ 408, L77

\bibitem{}  Baugh, C.M., Efstathiou, G. 1993, MNRAS 265, 145

\bibitem{}  Bond, J. R., Efstathiou, G. 1987, MNRAS, 226, 407

\bibitem{}  Clowes, R., Savage, A., Wang, G., et al. 1987, MNRAS 229, 27P

\bibitem{}  Collins, C.A., Joseph, R.D., Robertson, N.A. 1986, Nat, 320, 506

\bibitem{}  Davis, M., Peebles, P. J. E. 1983, ApJ 267, 465

\bibitem{}  Davis, M., Summers, F.J., Schlegel, 1992, Nat 359, 393

\bibitem{}  Efstathiou, G., Bond, J.R., White, S. 1992, MNRAS 258, 1p

\bibitem{}  Einasto, J., Gramann, M., Saar, E., et al. 1993, MNRAS 260, 705

\bibitem{}  Fabbri, R. 1988, ApJ, 334, 6

\bibitem{}  Fabbri, R. 1992, A\&A, 259, 1

\bibitem{}  Fabbri, R., Natale, V. 1990, ApJ, 363, 3

\bibitem{}  Fabbri, R., Natale, V.: 1993, A\&A, 267, L15

\bibitem{}  Fabbri, R., Torres, S. 1994, Report FS\#941U

\bibitem{}  Feldman, H.A., Kaiser, N., Peacock, J.A. 1993, ApJ 426, 23

\bibitem{}  Fisher, K. B., Davis, M., Strauss, M. A., et al. 1993, ApJ 402,
42

\bibitem{}  Guzzo, L., Iovino, A., Chincarini, G., et al. 1991, ApJ 382, L5

\bibitem{}  Jing, Y. P., Valdarnini, R. 1993, ApJ 406, 6

\bibitem{}  Lahav, O., Nemirof, R.J., Piran, T. 1990, ApJ 350, 119

\bibitem{}  Lawrence, A., Walker, D., Rowan-Robinson, M., et al. 1986,
MNRAS, 219, 687

\bibitem{}  Lucy, L.B. 1974, AJ 79, 745

\bibitem{}  Press, W.H., Flannery, B.P., Teukolsky, S.A., et al. 1988,
Numerical Recipes, Cambridge University Press, Cambridge

\bibitem{}  Rowan-Robinson, M., Lawrence, A., Saunders, W., et al. 1990,
MNRAS, 247, 1

\bibitem{}  Saunders, W., Rowan-Robinson, M., Lawrence, et al. 1990, MNRAS,
242, 318

\bibitem{}  Saunders, W., Rowan-Robinson, M., Lawrence, A. 1992, MNRAS 258,
134

\bibitem{}  Scaramella, R., Vittorio, N. 1993, MNRAS 263, L17

\bibitem{}  Scharf, C.A., Lahav, O. 1993, MNRAS 264, 439

\bibitem{}  Strauss, M. A., Davis, M., Yahil, A., et al. 1990, ApJ 361, 49

\bibitem{}  Taylor, A.N., Rowan-Robinson, M. 1992, Nat 359, 396

\bibitem{}  Torres, S., Fabbri, R., Ruffini, R. 1994, A\&A (in press)

\bibitem{}  Villumsen, J.V., Strauss, M. A. 1987, ApJ, 322, 37

\bibitem{}  Vogeley, M.S., Park, C., Geller, M.J., et al. 1992, ApJ 391, L5

\bibitem{}  White, M., Krauss, L.M., Silk, J. 1993, ApJ 418, 535

\bibitem{}  Yahil, A., Walker, D., Rowan-Robinson, M. 1986, ApJ, 301, L1
\end{thebibliography}
\end{document}